\documentclass[prl,twocolumn,superscriptaddress,longbibliography]{revtex4-2}
\usepackage[toc]{appendix}
\usepackage{amssymb,amsmath,amsfonts}
\usepackage{graphicx}
\usepackage{color,xcolor}
\usepackage{hyperref}
\usepackage{bm}
\usepackage{multirow}
\usepackage{natbib}

\usepackage{comment}

\usepackage[normalem]{ulem}
\usepackage{comment}
\usepackage{soul}

\begin{document}



\title{Reconstruction of surface electron spectrum  and cyclotron motion in the CDW phase of Weyl semimetals}

\author{A. A. Grigoreva}

\affiliation{\hbox{Department of Physics, University of Washington, Seattle, Washington 98195, USA}}

\author{A. V. Andreev}

\affiliation{\hbox{Department of Physics, University of Washington, Seattle, Washington 98195, USA}}

\author{L. I. Glazman} 
\affiliation{\hbox{Department of Physics, Yale University, New Haven, Connecticut 06520, USA}}

\date{May 7, 2024}

\begin{abstract}
Charge density wave (CDW) instability drastically affects the surface electron spectrum of a Weyl semimetal.
We show that in the CDW phase, the Fermi arcs reconnect into either closed Fermi loops or Frieze patterns traversing the reconstructed surface mini-Brillouin zone.  For the closed reconnection topology,
application of an out of plane magnetic field leads to a cyclotron motion of the surface electrons. We determine the cyclotron frequency as a function of the electron energy and the magnitude of the CDW gap $\Delta$ for various orientations of the Fermi arcs. For weak coupling, the period of cyclotron motion is dominated by the time of traversal of the arc reconnection regions and is inversely proportional to $\Delta$. 
\end{abstract}

\maketitle

Semimetals may undergo a transition to an excitonic insulator phase at low temperatures~\cite{Keldysh,CLOIZEAUX,Kohn}. This leads to a reconstruction of the electron spectrum, producing a gap at the Fermi level. In topological semimetals, the
Berry phase topology of the electron bands imposes constraints on the possible symmetry of exciton pairing. In particular, in Weyl semimetals (WSM)~\cite{Mele,Vafek_2014} exciton pairing between electrons and holes belonging to Weyl nodes with opposite chirality is possible only for the states with even angular momentum, $l=0,2,\ldots$, whereas pairing between nodes with the same chirality can occur only for odd $l$. This is a consequence of the quantized Berry monopole charge at the Weyl nodes~\cite{Haldane}. 

In this article, we consider the effect of the topological nature of the parent WSM state on the spectrum of surface states reconstructed by the excitonic transition in the bulk. We focus on time-reversal-invariant (TR) WSM, in which the exciton instability occurs between Weyl nodes with the opposite chirality and corresponds to $s$-wave pairing with a fully gapped Fermi surface.
Evidence for the realization of this instability in a TR-invariant WSM $(\mathrm{TaSe}_4)_2\mathrm{I}$ has been reported in Ref.~\cite{gooth2019axionic}, although that interpretation was subsequently challenged in Ref.~\cite{Sinchenko}. Signatures of an excitonic insulator phase were also seen \cite{sun2022evidence,fei2017edge} in the electron transport in a monolayer $\mathrm{WTe}_2$, whose bulk is believed to be a type-II WSM.

\begin{figure}[h!]
    \centering
    \includegraphics[width=0.40\textwidth]{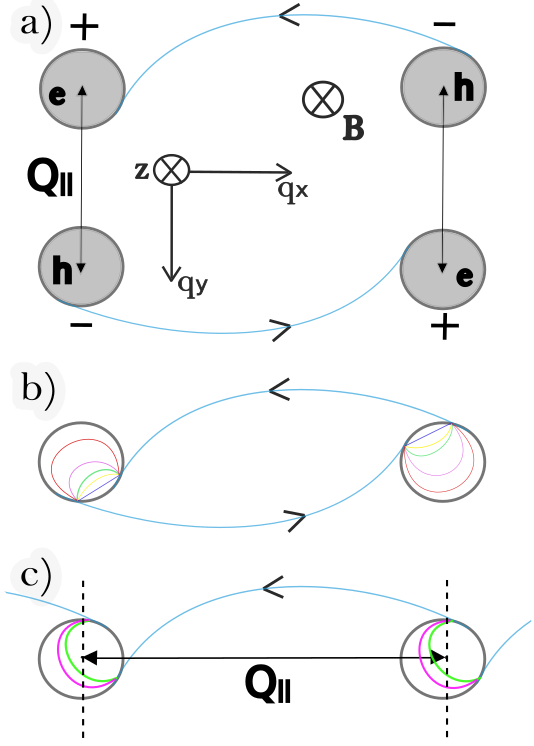}
    \caption{Panel \emph{a)}: Projection of arc states (blue lines) and bulk states (circles) at the Fermi level onto the momentum plane parallel to the surface. The node chirality is indicated by ``+" and ``-". 
    Panel \emph{b)}: Closed Fermi arc reconnection caused by CDW formation after zone folding.  Different colors correspond to different energies. Panel \emph{c)}: In the open reconnection topology,  the reconstructed spectrum forms a Frieze pattern in the extended Brillouin zone picture. Only the Frieze pattern for the top pair of Weyl nodes is shown. The pattern for the bottom pair is related to it by TR symmetry. 
    The dashed lines indicate the boundaries of the surface Brillouin mini-zone.  }
  \label{fig:4}
\end{figure}

The reconstruction of the surface state spectrum in the excitonic transition is illustrated in Fig.~\ref{fig:4}. In the WSM phase (panel \emph{a)}), the spectrum consists of the Fermi arcs that merge with the Fermi circles of bulk states, which are the projections of the Fermi surfaces (hereinafter assumed spherical),  of the bulk spectrum onto the surface momentum plane. The excitonic transition corresponds to the formation of a charge density wave (CDW) in the bulk with a wave vector $\bm{Q}$ that connects the centers of the corresponding Weyl nodes.
The reconstructed surface Brillouin zone (SBZ) is obtained by folding the semimetal  SBZ by the surface projection $\bm{Q}_\parallel$ of the CDW wavevector $\bm{Q}$, as indicated in panels \emph{b)} and \emph{c)}.  
The constant energy lines of the surface states consist of the Fermi arcs of the original WSM, which are connected by newly formed arcs inside the Fermi circles. There are two possible reconnection topologies: closed loops illustrated in panel \emph{b)}, or Frieze patterns spanning the folded SBZ, as shown in panel \emph{c)}.

A closed Fermi arc reconnection topology allows for the conventional cyclotron motion in the presence of an out-of-plane magnetic field. Alternatively, in the case of the Frieze pattern, electron velocity varies periodically in time, while the real-space ``cyclotron" orbits are open. At small values of the exciton gap $\Delta$ the period of the cyclotron motion is dominated by the 
traversal time of the segments connecting the arcs, see Fig.~\ref{fig:4} b), c).

We consider the minimal model of a TR-invariant WSM with two electron-populated Weyl nodes related by TR, and two TR-related hole-populated nodes. By TR symmetry, the two electron nodes must have the same chirality, and thus by the Nielsen-Ninomiya theorem~\cite{nielsen1981absence1,nielsen1981absence2} the two hole nodes must have the opposite chirality. As a result, the excitonic instability occurs between the opposite chirality nodes and is assumed to correspond to $s$-wave exciton pairing~\cite{Wei2012}.

For simplicity, we assume that the electron dispersion in the vicinity of the Weyl nodes is isotropic, and the band velocity is the same in the electron- and hole- nodes. In a compensated system, the electron- and hole- Fermi surfaces are congruent spheres of radius $k_F$ centered at the appropriate Weyl nodes. Below, we use dimensionless variables; the momenta $\bm{p}$ are measured in units of the Fermi momentum $k_F$, and energies - in units of the Fermi energy $\epsilon_F$. We also measure energies from the chemical potential (the Weyl nodes in the electron/hole-populated valleys are offset by $\pm 1$ from the chemical potential).

Since the electron and hole Fermi circles are congruent, the semimetal state becomes unstable with respect to the exciton condensate formation at zero temperature for arbitrarily weak repulsion~\cite{Keldysh}. We assume the repulsive interaction to be weak and describe the spectral reconstruction using the mean-field theory. The indirect exciton pairing between electron- and hole- Weyl nodes with opposite chirality
is described by a CDW potential with a wave vector $\bm{Q}$ which connects the two nodes. Upon Brillouin zone folding by $\bm{Q}$, the electron  spectrum in the excitonic phase is described by the eigenvalues of the following $4\times 4$ matrix Hamiltonian,
\begin{equation}
    \left[(\bm{\sigma} \bm{p} - 1) \widehat \tau_{3} + \Delta \widehat \tau_{1}\right]\bm{\Psi} = \varepsilon \bm{\Psi}.
    \label{eq:SurfHam}
\end{equation} 
Here $\Delta \ll 1$ is the dimensionless pairing gap, and we introduced the $2 \times 2$ Pauli matrices $\sigma$ and $\tau$. The $\sigma$ matrices describe the Weyl spectrum in each node, and the $\tau$-matrices  act in the space of nodes.  In these notations the four-component spinor $\bm{\Psi}$ wavefunction is given by
\begin{equation}
           \bm{\Psi}=\begin{pmatrix}\bm{u}\\
\bm{w}
\end{pmatrix}, \quad \bm{u}=\begin{pmatrix}u\\
v
\end{pmatrix},\quad \bm{w}=\begin{pmatrix}w\\
h
\end{pmatrix} .
\label{eq:WF}
\end{equation}
where the two-component spinors $\bm{u}$ and $\bm{w}$ correspond to the electron- and hole nodes, respectively.

We assume that the crystal surface is normal to the $z$-axis. Therefore,  we express the momentum operator in Eq.~\eqref{eq:SurfHam} as $\bm{p} = (q_{x}, q_{y}, -i \partial_{z})$, with $\bm{q}= (q_x,q_y)$ being  the conserved momentum along the surface. This turns Eq.~\eqref{eq:SurfHam} into  a first-order matrix differential equation whose solution is given by 
\begin{align}\label{eq:Psi_kappa}
    \bm{\Psi}_{\bm{q}}(z) = e^{-\widehat{ \kappa}(\varepsilon,\bm{q}) z} \bm{\Psi}_{\bm{q}}(0), 
    \quad  \bm{\Psi}_{\bm{q}}(0) = \begin{pmatrix}
    u_{\bm{q}}\\
v_{\bm{q}}\\
w_{\bm{q}}\\
h_{\bm{q}}
\end{pmatrix},
\end{align}
where the $4 \times 4$ matrix $\widehat{\kappa}(\varepsilon,
\bm{q})$ is given by  Eq.~(SM.2) in the Supplemental material \cite{supp}.

The equations above must be supplemented by the boundary conditions. We consider the case of weak coupling, $\Delta \ll 1$, and assume that the boundary conditions are unaffected by the CDW pairing. In this case, the boundary conditions do not mix the spinor amplitudes $\bm{u}$ and $\bm{w}$, and correspond to the vanishing of the $z$-component of current in each node, $\bm{u}^\dagger \sigma_z \bm{u}=\bm{w}^\dagger \sigma_z \bm{w} =0 $. In this case, the most general boundary condition is given by~\footnote{We consider the simplest situation, where the band-bending effects near the surface are neglected. In the WSM phase band-bending creates a spiraling structure of the Fermi arcs~\cite{li2015spiraling}. The influence of band-bending on the surface spectrum in the CDW phase is outside the scope of the present work.  }
\begin{equation}
    \begin{pmatrix}
    u_{\bm{q}}\\
w_{\bm{q}}
\end{pmatrix} = \begin{pmatrix} e^{i\chi_e} & 0\\
0 & e^{i\chi_h}
\end{pmatrix} \begin{pmatrix}
v_{\bm{q}}\\
h_{\bm{q}}
\end{pmatrix}.
\label{eq:BC}
\end{equation}
The phases $\chi_{e}$ and $\chi_{h}$ determine the orientation of the Fermi arcs. For our choice of right-handed electron node and left-handed hole node in Eq.~\eqref{eq:SurfHam}, the electron (hole) arcs touch the Fermi circle at angles $- \chi_{e}$ ($-\chi_{h}$) and emanate in the counterclockwise direction, as shown in Fig.~\ref{fig:5}. Below, without loss of generality, we set $ \chi_{h} = + \chi$ and $\chi_{e} = -\chi$. 
It is clear from Eqs.~\eqref{eq:SurfHam} and \eqref{eq:WF} 
that this can be achieved by a frame rotation in the $q_x-q_y$ plane.

\begin{figure}[h!]
    \centering
\includegraphics[width=0.45\textwidth]{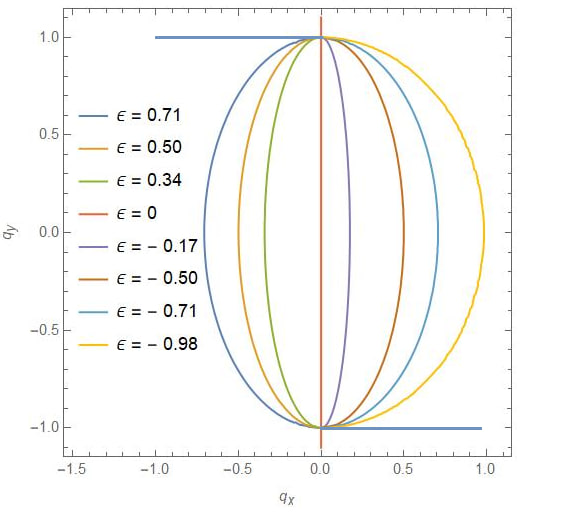}
    \caption{Constant energy cuts of spectrum given by Eq. \eqref{eq:MainEq} for different values of energy. This plot for the value $\chi = \frac{\pi}{2}$. The straight horizontal lines show the Fermi arcs.}
  \label{fig:Figure3}
\end{figure}

For weak coupling, $\Delta \ll 1$,  the surface spectrum acquires a particularly simple form for momenta $\bm{q}$ outside  narrow annular regions of width $\sim \Delta$ near the respective Fermi circles, $||\bm{q}|-1| \gg \Delta$.  In this case, outside the circles the spectrum and the wavefunctions coincide with those of the original Fermi arcs of the WSM. Inside the Fermi circles, new surface states appear. We now summarize the main steps of the derivation of the spectrum of these states,  delegating the details to Supplemental Material (SM). 
The surface states  correspond to the solutions that decay exponentially into the bulk. Therefore, they must be given by a linear combination of the two eigenvectors of  $\widehat{\kappa}(\varepsilon,
\bm{q})$, which correspond to the eigenvalues with a positive real part, see Eqs. (SM.3) and (SM.4). Requiring that the linear combination satisfies the boundary condition \eqref{eq:BC} we find the surface states spectrum 
 Eq.~(SM.7), which for $1-|\bm{q}| \gg \Delta$  simplifies to
\begin{align}
    c - q_{x} =\epsilon s \sqrt{\frac{1 - q^{2}_{y} - q^{2}_{x}}{1-\epsilon^2 }}; \quad c\equiv \cos \chi, \quad s \equiv  \sin \chi.
    \label{eq:MainEq}
\end{align}
Here we introduced a new dimensionless energy measured in units of $\Delta$, $\epsilon = \varepsilon/\Delta$, and the shorthand notations $c$ and $s$.

For energies inside the gap, $|\epsilon |< 1$,  Eq.~\eqref{eq:MainEq} describes arcs of an ellipse centered 
at $q_x =  \frac{c(1-\epsilon^2)}{1-\epsilon^2 c^2}$, and  $  q_y =0$, and the semi-axes along $q_x$ and $q_y$ given by, respectively, $a = \frac{|\epsilon | s^2}{1 - \epsilon^2 c^2}$ and $b = \frac{s}{\sqrt{1 - c^2\epsilon^2}}$. The ellipse is inscribed in the Fermi circle and touches it at points  $\bm{q}_\pm = (q_x,q_y)=(c,\pm s)$.  The choice of the arc is dictated by the sign of $\epsilon$. For positive energies the arcs lie on one side of the straight chord $q_x = \cos \chi$, and for negative energies - on the other. This is shown in Fig.~\ref{fig:Figure3} for $\chi = \pi/2$. At $\epsilon=0$ the elliptic arc degenerates into the straight chord $q_x = \cos \chi$. When the energy approaches the top and bottom of the band gap, $\epsilon \to \pm \Delta$, the elliptic arc approaches one of the two arcs of the unit Fermi circle, which connect the points $\bm{q}_\pm$. 
The oriented area of the surface confined between the elliptic arc and the chord (in dimensionless units) is given by
\begin{align}
\label{eq:dimensionless_area}
    \mathcal{A}(\epsilon) 
    = & \,\frac{\epsilon s^3 }{\left( 1- c^2 \epsilon^2 \right)^{3/2} } \left(\arccos (c \epsilon)  - c \epsilon \sqrt{1 -c^2 \epsilon^2}  \right) .
\end{align}
As the energy increases across the gap, the oriented area increases monotonically. For opposite signs of $\epsilon$, the oriented area $\mathcal{A}(\epsilon)$  is shown by different colors in Fig.~\ref{fig:5}. 

\begin{figure}[h!]
    \centering
\includegraphics[width=0.45\textwidth]{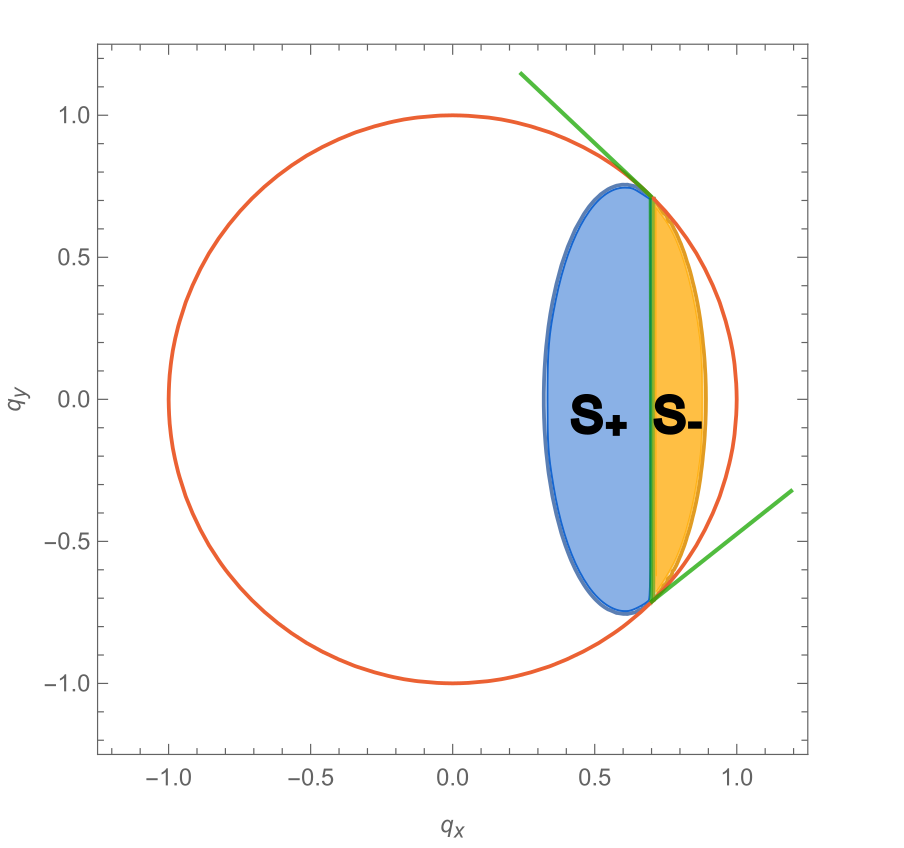}
    \caption{Oriented areas $S_\pm = \mathcal{A}(\pm 1/2)$, confined between the chord and the elliptic arcs in Eq.~\eqref{eq:dimensionless_area} for $\epsilon = \pm 1/2$  and $\chi = \frac{\pi}{4}$.}
  \label{fig:5}
\end{figure}

Since the original Fermi arcs touch the unit circle precisely at points $\bm{q}_\pm$ we see that the new elliptic arcs that appear due to spectrum reconstruction connect the original Fermi arcs that terminate at the Fermi circle. The explicit formulas, which describe the fine structure of this reconnection inside the narrow strip $|q-1|\lesssim \Delta^2$ become rather cumbersome. Since they are not important for the general picture of spectral reconstruction we do not present them here.

\emph{Surface cyclotron motion:} Next, we discuss the implications of the reconstruction of the surface state spectrum determined above for the surface cyclotron motion in the excitonic phase. We consider the geometry in which the magnetic field is normal to the surface. In the WSM phase, in a slab geometry an unusual cyclotron motion was proposed in Ref.~\cite{Potter2014,moll2016transport}. In this motion, the cyclotron trajectory consists of segments of motion along the Fermi arc states on the opposite surfaces of the slab, which are connected by the chiral cyclotron motion through the bulk of the slab.

In the excitonic phase, the motion through the bulk becomes impossible, and the electrons at opposite surfaces of the slab become decoupled. The system becomes a topological insulator (TI). Depending on the arc reconnection topology, the momentum-space trajectories of surface electrons are either closed loops or open curves. For the closed topology of Fermi arc reconnection a surface cyclotron motion becomes possible, which is similar to the one in TIs, observed in
Ref.~\cite{Analytis2010}.

Using the above results for the reconstructed surface state spectrum, the surface cyclotron motion can be considered using the standard methods~\cite{Abrikosov-book,Ashcroft}.  The momentum-space trajectory corresponds to the constant energy curve, whereas the real space trajectory is obtained by rotating the one in momentum-space by $\pi/2$ and rescaling it by $l_H^2= \hbar c /eH$. For the closed arc reconnection topology of Fig.~\ref{fig:4} b) the period of the cyclotron motion is given by the standard expression,\begin{equation}
    T(E) = l^{2}_{H} \frac{d S (E)}{d E},
    \label{eq:Time_Schockley}
\end{equation}
where $S(E)$ is the momentum space area confined by the curve of constant energy $E$. 

It is instructive to separate the cyclotron period in Eq.~\eqref{eq:Time_Schockley} into the time of traversal of the Fermi arcs, $t_{\mathrm{arc}}$, and of the arc reconnection regions inside the gapped Fermi circles, $t_{\mathrm{rec}} (\epsilon)$,
\begin{equation}
\label{eq:T_decomposition}
    T (\epsilon) = 2 t_{\mathrm{rec}} (\epsilon) + 2t_{\mathrm{arc}}. 
\end{equation}
Here we used the fact that in a TR-invariant system, the times of traversal of the two arcs and the two reconnection regions are pairwise equal. 
At weak CDW pairing $t_{\mathrm{arc}}$ is practically independent of $\Delta$ and the particle energy $\epsilon$. In contrast, $t_{\mathrm{rec}} (\epsilon)$ is inversely proportional to $\Delta$ and exhibits a strong energy dependence.

\begin{figure}[h!]
    \centering
\includegraphics[width=0.45\textwidth]{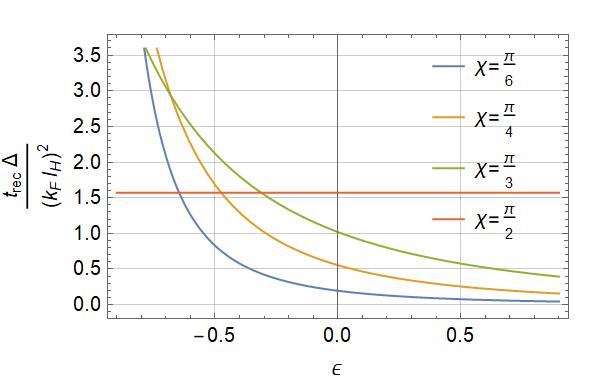}
    \caption{Time of traversal of the reconnection arc, $t_{\mathrm{rec}}(\epsilon, \chi)$ in Eq.~\eqref{eq:traversal_time}  as a function of energy for different values of the angle $\chi$. Because of the symmetry  $t_{\mathrm{rec}}(\epsilon, \pi/2 -\chi) = t_{\mathrm{rec}}(- \epsilon, \chi)$ only graphs for $\chi \in (0, \pi/2)$ are presented. Units on the vertical are chosen to be $t_{\mathrm{rec}}(\epsilon) \Delta/(k_F l_H)^2$ for convenience.} 
  \label{fig:FigureTime}
\end{figure}

The decomposition of the cyclotron period in Eq.~\eqref{eq:Time_Schockley} into the two contributions in Eq.~\eqref{eq:T_decomposition} is accomplished by representing the momentum space area enclosed by the constant energy line as a sum of the area confined by the straight chords and the Fermi arcs, $S_{\mathrm{arc}}(\epsilon)$, and the oriented area $k_F^2 \mathcal{A}(\epsilon)$ confined between the chords and the elliptic arcs. The latter is described by Eq.~\eqref{eq:dimensionless_area} and shown in Fig.~\ref{fig:5}. This yields, $t_{\mathrm{rec}}(\epsilon) = k_F^2l_H^2 \partial_\epsilon \mathcal{A}(\epsilon)/\Delta$. Using Eq.~\eqref{eq:dimensionless_area} we get
\begin{align}
\label{eq:traversal_time}
    t_{\mathrm{rec}} (\epsilon)  = & \, \frac{k_F^2 l_H^2 s^3 }{\Delta}  \frac{\arccos(c\epsilon) (1+ 2 c^2 \epsilon^2)  - 3 \epsilon c \sqrt{1-c^2 \epsilon^2}}{(1-c^2 \epsilon^2)^{5/2}}. 
\end{align}
This function is plotted in Fig.~\eqref{fig:FigureTime} for several values of $\chi$.
The positive sign of $t(\epsilon)$, leading to $T(E)>0$ for the cyclotron motion is ``inherited" from the bulk WSM properties. It corresponds to the evolution from hole- to particle-dominated content of the  charge carriers upon the increase of energy.
It is clear that a simultaneous change $\epsilon \to - \epsilon$, $\chi \to \pi/2 - \chi$ leaves $t_{\mathrm{rec}} (\epsilon)$ invariant. The sign of the derivative, $dt/d\epsilon$, is not universal and depends on $\chi$. In particular, for $\chi=\pi/2$, the derivative vanishes. In this case   $t_{\mathrm{rec}} (\epsilon)$ is independent of energy and is equal to $\pi k_F^2 l_H^2/2 \Delta$.

Thus we conclude that at $\Delta\ll 1$ the period of cyclotron motion is dominated by the time of traversal of the arc reconnection region, $|q|<1$. In this region, electrons undergo a bouncing motion along the $z$-axis within a length $ \xi \sim v/\Delta$ (with $v$ being the node velocity) near the surface of the crystal, combined with the drift parallel to the surface.  For $\Delta \ll 1$ the bouncing period becomes long, resulting in a small cyclotron frequency.

For the open reconnection topology shown in Fig.~\ref{fig:Figure3} \emph{b)},  the constant energy contours in momentum space correspond to pairs of open curves that span the reconstructed surface Brillouin mini-zone. In this case, the position space cyclotron orbits are open and correspond to the undulating motion. The undulation period corresponds to the time of traversal of the Brillouin mini-zone, and is given by half of  $T(\epsilon)$ in Eq.~\eqref{eq:T_decomposition}.

In summary, we elucidated the transformation of a Weyl semimetal  into a topological insulator, resulting from the exciton instability in the bulk of WSM. The gap opening in the electron spectrum in the bulk of WSM is accompanied by a reconnection of the Fermi arcs and formation of two-dimensional Fermi seas at the surfaces, which is characteristic of a TI. An out-of-plane magnetic field induces electron cyclotron motion separate for the opposite surfaces, contrary to the respective effect in a conventional WSM.

We thank Jiun-Haw Chu, David Cobden, Joel Moore, Andrew Potter, and Dmitry Zverevich for useful discussions. 
This work was supported by the National Science Foundation through the MRSEC Grant No. DMR-1719797, the Thouless Institute for Quantum Matter (TIQM), and the College of Arts and Sciences at the University of Washington (A.G. and A.A.), and by Office of Naval Research (ONR) under Award No. N00014-22-1-2764 (L.I.G.).

\appendix

\section{Supplemental Material}

Here we derive Eq.~(5), which gives a simplified expression  for the surface electron spectrum in the interior of the Fermi circle for $\Delta \ll 1$ and   $1-|q|\gg \Delta$.

Writing the momentum  in Eq.~(1) in the form $\bm{p}  =(q_{x}, q_{y}, -i \partial_{z})$  we obtain 
\begin{equation} \tag{SM.1}
\label{eq:ODE_z}
    \partial_{z}\bm{\Psi}_{\bm{q}} = - \widehat \kappa(\varepsilon,\bm{q}) \bm{\Psi}_{\bm{q}},
\end{equation}
where $\bm{q} =(q_{x},q_{y})$ and   $\widehat \kappa (\varepsilon, \bm{q})$ is a $4\times 4$ matrix defined as
\begin{equation} \tag{SM.2}
    \widehat \kappa(\varepsilon, \bm{q}) = \begin{pmatrix}-i\left(\varepsilon+1\right) & iq^{*} & i\Delta & 0\\
-iq & i\left(\varepsilon+1\right) & 0 & -i\Delta\\
-i\Delta & 0 & i\left(\varepsilon-1\right) & iq^{*}\\
0 & i\Delta & -iq & -i\left(\varepsilon-1\right)
\end{pmatrix}.
\label{eq:Kappa}
\end{equation}
Here we introduced the complex variables for the in-plane momentum, $q = q_{x} + i q_{y}$, and $q^* = q_{x} - i q_{y} $.

Since we are interested in energies below the gap it is more convenient to work with the rescaled energy $\epsilon= \varepsilon/\Delta$. The eigenvalues of $\widehat{\kappa}$ form two complex-conjugate pairs. The pair with a positive real part is given by
\begin{align} \tag{SM.3}
    \label{eq:lambda_pm}
    \lambda_\pm = & \, \pm i \sqrt{1-|q|^2 -\Delta^2 (1-\epsilon^2)  \mp 2 i \Delta  \sqrt{1-\epsilon^2}},
\end{align}
where the branches of the square roots are chosen to have zero argument on the positive real axis of their variables (the pair of the eigenvalues with a negative real part is given by $-\lambda_\pm$).
The eigenvectors corresponding to the eigenvalues $\lambda_\pm$ in Eq.~\eqref{eq:lambda_pm} are given by
\begin{align} \tag{SM.4}
\label{eq:eigenvectors}
    V_\pm = &\, 
    \begin{pmatrix}
        u_\pm \\
        v_\pm \\
        w_\pm \\
        h_\pm
    \end{pmatrix} =
    \begin{pmatrix}   \left( 1 \mp i \Delta \sqrt{1 - \epsilon^2} + i\lambda_{\pm}\right) \left(\epsilon \mp  i \sqrt{1 - \epsilon^{2}}\right) \\
    q \left(\epsilon \mp i \sqrt{1-\epsilon^2}\right)\\
    1 \mp i\Delta\sqrt{1-\epsilon^2} + i\lambda_{\pm }\\
    q
    \end{pmatrix}.
\end{align}
Note that $\frac{u_\pm}{v_\pm} =\frac{w_\pm}{h_\pm}$. This is a manifestation of the fact that 
the energy eigenstates $V_\pm$ are also eigenstates of helicity, $\bm{p}\cdot \bm{\sigma}/|\bm{p}|$ (the helicity eigenvalues are different for $v_\pm$).

For surface states, the linear combination $V_+ + rV_-$, with $r$ being a complex number, must satisfy the boundary condition Eq.~(4). 
Substitution of  $\bm{\Psi}_{\bm{q}}(0) = V_+ + r V_-$, into Eq.~(4) with $\chi_\pm = \pm \chi$, yields the following conditions 
\begin{align} \tag{SM.5}
\frac{u_+ + r u_-}{v_+ + r v_-} = & e^{i \chi} , \quad 
\frac{w_+ + r w_-}{h_+ + r h_-} =  e^{-i \chi}.   
\end{align} 
Excluding $r$ we obtain
\begin{align}
  & e^{i \chi} \left(v_- w_+ - v_+ w_-\right) + e^{- i \chi} \left(u_- h_+ - u_+ h_-\right) = \nonumber \\ 
  & u_- w_+ - u_+ w_- + v_- h_+ - v_+ h_-.  \tag{SM.6}
\end{align} 
Substituting Eq.~\eqref{eq:eigenvectors} here we get
\begin{align} 
       & \left(1-i\Delta \sqrt{1 - \epsilon^2} + i\lambda_{+}\right) \left(1+i\Delta \sqrt{1 - \epsilon^2} + i\lambda_{-}\right)  =   \nonumber \\
    &  - q^2 + q e^{i\chi} \left[1 -  \Delta \epsilon  + \frac{ \epsilon (\lambda_+ - \lambda_-) }{ \sqrt{1 - \epsilon^2}} + \frac{i}{2} (\lambda_+ + \lambda_-)\right] \nonumber \\
     & + q e^{- i \chi} \left[1 +  \Delta \epsilon  - \frac{ \epsilon (\lambda_+ - \lambda_-) }{ \sqrt{1 - \epsilon^2}} + \frac{i}{2} (\lambda_+ + \lambda_-)\right].  \tag{SM.7}
\label{eq:Eq1}
\end{align}

In the interior of the Fermi circle,
$1 - |q|^{2} \gg \Delta$, the eigenvalues $\lambda_\pm$ in Eq.~\eqref{eq:lambda_pm}  may be approximated as 
\begin{align}  \tag{SM.8}
    \label{eq:lambda_approx}
    \lambda_{\pm} \approx  \sqrt{1 - |q|^2} \left( \pm i  + \frac{\Delta}{1-|q|^2} \sqrt{  1 - \epsilon^{2}}\right).
\end{align}
Substituting this approximation into  Eq.~\eqref{eq:Eq1} and neglecting the  terms that are small in we obtain the approximate equation for the spectrum, applicable  $1-|q|\gg \Delta$ 
\begin{align}
    \label{eq:approximate_inside}
      |q|^2 + q^2  = &\,   2 q  \left[ \cos \chi - \sin \chi \frac{\epsilon \sqrt{1-|q|^2}}{\sqrt{1-\epsilon^2}} \right].   \tag{SM.9}
\end{align}
In Cartesian coordinates $q_x, q_y$ this yields,
\[
2 q_x(q_x + i q_y) = 2 (q_x + i q_y)\left[ \cos \chi - \sin \chi \frac{\epsilon \sqrt{1-|q|^2}}{\sqrt{1-\epsilon^2}} \right].   \tag{SM.10}
\]
This reproduces Eq.~(7) in the main text. 

Outside the Fermi circle, for 
$ |q|^{2} -1 \gg \Delta$, the eigenvalues $\lambda_\pm$ in Eq.~\eqref{eq:lambda_pm}  may be approximated as 
\begin{align}
\label{eq:lambda_approx_outside}
    \lambda_{\pm} \approx \lambda = \sqrt{ |q|^2 -1 } .  \tag{SM.11}
\end{align}
Therefore, setting $\lambda_+ - \lambda_-\to 0$ in Eq.~\eqref{eq:Eq1} and neglecting terms with $\Delta$, we get 
\begin{align}
    \label{eq:outside}
    1+i \lambda = & \, q e^{\pm i \chi}.  \tag{SM.12}
\end{align}
It is convenient to introduce the angle $\theta$ formed by the vector $\bm{q}$ and the radius to the point of contact between the arc the Fermi circle. Since the Fermi circle has unit radius in our units, we have 
$1/|q|=\cos\theta$. Substituting this into (SM.11) and using (SM.10) we get
\[
|q| e^{i \theta }= q e^{\pm i \chi}, \quad e^{i \theta }= \frac{q}{|q|} e^{\pm i \chi}.  \tag{SM.13}
\]
This reproduces the  original Fermi arcs in the WSM phase. 

\bibliography{WSM}

\begin{thebibliography}{20}%
\makeatletter
\providecommand \@ifxundefined [1]{%
 \@ifx{#1\undefined}
}%
\providecommand \@ifnum [1]{%
 \ifnum #1\expandafter \@firstoftwo
 \else \expandafter \@secondoftwo
 \fi
}%
\providecommand \@ifx [1]{%
 \ifx #1\expandafter \@firstoftwo
 \else \expandafter \@secondoftwo
 \fi
}%
\providecommand \natexlab [1]{#1}%
\providecommand \enquote  [1]{``#1''}%
\providecommand \bibnamefont  [1]{#1}%
\providecommand \bibfnamefont [1]{#1}%
\providecommand \citenamefont [1]{#1}%
\providecommand \href@noop [0]{\@secondoftwo}%
\providecommand \href [0]{\begingroup \@sanitize@url \@href}%
\providecommand \@href[1]{\@@startlink{#1}\@@href}%
\providecommand \@@href[1]{\endgroup#1\@@endlink}%
\providecommand \@sanitize@url [0]{\catcode `\\12\catcode `\$12\catcode `\&12\catcode `\#12\catcode `\^12\catcode `\_12\catcode `\%12\relax}%
\providecommand \@@startlink[1]{}%
\providecommand \@@endlink[0]{}%
\providecommand \url  [0]{\begingroup\@sanitize@url \@url }%
\providecommand \@url [1]{\endgroup\@href {#1}{\urlprefix }}%
\providecommand \urlprefix  [0]{URL }%
\providecommand \Eprint [0]{\href }%
\providecommand \doibase [0]{https://doi.org/}%
\providecommand \selectlanguage [0]{\@gobble}%
\providecommand \bibinfo  [0]{\@secondoftwo}%
\providecommand \bibfield  [0]{\@secondoftwo}%
\providecommand \translation [1]{[#1]}%
\providecommand \BibitemOpen [0]{}%
\providecommand \bibitemStop [0]{}%
\providecommand \bibitemNoStop [0]{.\EOS\space}%
\providecommand \EOS [0]{\spacefactor3000\relax}%
\providecommand \BibitemShut  [1]{\csname bibitem#1\endcsname}%
\let\auto@bib@innerbib\@empty
\bibitem [{\citenamefont {Keldysh}\ and\ \citenamefont {Kopaev}(1965)}]{Keldysh}%
  \BibitemOpen
  \bibfield  {author} {\bibinfo {author} {\bibfnamefont {L.~V.}\ \bibnamefont {Keldysh}}\ and\ \bibinfo {author} {\bibfnamefont {Y.~V.}\ \bibnamefont {Kopaev}},\ }\bibfield  {title} {\bibinfo {title} {Possible instability of the semimetallic state toward coulomb interaction},\ }\href@noop {} {\bibfield  {journal} {\bibinfo  {journal} {Sov. Phys. Solid State}\ }\textbf {\bibinfo {volume} {6}},\ \bibinfo {pages} {2219} (\bibinfo {year} {1965})}\BibitemShut {NoStop}%
\bibitem [{\citenamefont {Cloizeaux}(1965)}]{CLOIZEAUX}%
  \BibitemOpen
  \bibfield  {author} {\bibinfo {author} {\bibfnamefont {J.~D.}\ \bibnamefont {Cloizeaux}},\ }\bibfield  {title} {\bibinfo {title} {Exciton instability and crystallographic anomalies in semiconductors},\ }\href {https://doi.org/https://doi.org/10.1016/0022-3697(65)90153-8} {\bibfield  {journal} {\bibinfo  {journal} {Journal of Physics and Chemistry of Solids}\ }\textbf {\bibinfo {volume} {26}},\ \bibinfo {pages} {259} (\bibinfo {year} {1965})}\BibitemShut {NoStop}%
\bibitem [{\citenamefont {J\'erome}\ \emph {et~al.}(1967)\citenamefont {J\'erome}, \citenamefont {Rice},\ and\ \citenamefont {Kohn}}]{Kohn}%
  \BibitemOpen
  \bibfield  {author} {\bibinfo {author} {\bibfnamefont {D.}~\bibnamefont {J\'erome}}, \bibinfo {author} {\bibfnamefont {T.~M.}\ \bibnamefont {Rice}},\ and\ \bibinfo {author} {\bibfnamefont {W.}~\bibnamefont {Kohn}},\ }\bibfield  {title} {\bibinfo {title} {Excitonic insulator},\ }\href {https://doi.org/10.1103/PhysRev.158.462} {\bibfield  {journal} {\bibinfo  {journal} {Phys. Rev.}\ }\textbf {\bibinfo {volume} {158}},\ \bibinfo {pages} {462} (\bibinfo {year} {1967})}\BibitemShut {NoStop}%
\bibitem [{\citenamefont {Armitage}\ \emph {et~al.}(2018)\citenamefont {Armitage}, \citenamefont {Mele},\ and\ \citenamefont {Vishwanath}}]{Mele}%
  \BibitemOpen
  \bibfield  {author} {\bibinfo {author} {\bibfnamefont {N.~P.}\ \bibnamefont {Armitage}}, \bibinfo {author} {\bibfnamefont {E.~J.}\ \bibnamefont {Mele}},\ and\ \bibinfo {author} {\bibfnamefont {A.}~\bibnamefont {Vishwanath}},\ }\bibfield  {title} {\bibinfo {title} {Weyl and {Dirac} semimetals in three-dimensional solids},\ }\href {https://doi.org/10.1103/RevModPhys.90.015001} {\bibfield  {journal} {\bibinfo  {journal} {Rev. Mod. Phys.}\ }\textbf {\bibinfo {volume} {90}},\ \bibinfo {pages} {015001} (\bibinfo {year} {2018})}\BibitemShut {NoStop}%
\bibitem [{\citenamefont {Vafek}\ and\ \citenamefont {Vishwanath}(2014)}]{Vafek_2014}%
  \BibitemOpen
  \bibfield  {author} {\bibinfo {author} {\bibfnamefont {O.}~\bibnamefont {Vafek}}\ and\ \bibinfo {author} {\bibfnamefont {A.}~\bibnamefont {Vishwanath}},\ }\bibfield  {title} {\bibinfo {title} {{Dirac} fermions in solids: From high-{$T_{c}$} {Cuprates} and {Graphene} to {Topological Insulators} and {Weyl Semimetals}},\ }\href {https://doi.org/10.1146/annurev-conmatphys-031113-133841} {\bibfield  {journal} {\bibinfo  {journal} {Annual Review of Condensed Matter Physics}\ }\textbf {\bibinfo {volume} {5}},\ \bibinfo {pages} {83} (\bibinfo {year} {2014})}\BibitemShut {NoStop}%
\bibitem [{\citenamefont {Li}\ and\ \citenamefont {Haldane}(2018)}]{Haldane}%
  \BibitemOpen
  \bibfield  {author} {\bibinfo {author} {\bibfnamefont {Y.}~\bibnamefont {Li}}\ and\ \bibinfo {author} {\bibfnamefont {F.~D.~M.}\ \bibnamefont {Haldane}},\ }\bibfield  {title} {\bibinfo {title} {Topological nodal cooper pairing in doped weyl metals},\ }\href {https://doi.org/10.1103/PhysRevLett.120.067003} {\bibfield  {journal} {\bibinfo  {journal} {Phys. Rev. Lett.}\ }\textbf {\bibinfo {volume} {120}},\ \bibinfo {pages} {067003} (\bibinfo {year} {2018})}\BibitemShut {NoStop}%
\bibitem [{\citenamefont {Gooth}\ \emph {et~al.}(2019)\citenamefont {Gooth}, \citenamefont {Bradlyn}, \citenamefont {Honnali}, \citenamefont {Schindler}, \citenamefont {Kumar}, \citenamefont {Noky}, \citenamefont {Qi}, \citenamefont {Shekhar}, \citenamefont {Sun}, \citenamefont {Wang} \emph {et~al.}}]{gooth2019axionic}%
  \BibitemOpen
  \bibfield  {author} {\bibinfo {author} {\bibfnamefont {J.}~\bibnamefont {Gooth}}, \bibinfo {author} {\bibfnamefont {B.}~\bibnamefont {Bradlyn}}, \bibinfo {author} {\bibfnamefont {S.}~\bibnamefont {Honnali}}, \bibinfo {author} {\bibfnamefont {C.}~\bibnamefont {Schindler}}, \bibinfo {author} {\bibfnamefont {N.}~\bibnamefont {Kumar}}, \bibinfo {author} {\bibfnamefont {J.}~\bibnamefont {Noky}}, \bibinfo {author} {\bibfnamefont {Y.}~\bibnamefont {Qi}}, \bibinfo {author} {\bibfnamefont {C.}~\bibnamefont {Shekhar}}, \bibinfo {author} {\bibfnamefont {Y.}~\bibnamefont {Sun}}, \bibinfo {author} {\bibfnamefont {Z.}~\bibnamefont {Wang}}, \emph {et~al.},\ }\bibfield  {title} {\bibinfo {title} {Axionic charge-density wave in the weyl semimetal {$(\mathrm{TaSe}_{4})_{2}\mathrm{I}$}},\ }\href {https://doi.org/10.1038/s41586-019-1630-4} {\bibfield  {journal} {\bibinfo  {journal} {Nature}\ }\textbf {\bibinfo {volume} {575}},\ \bibinfo {pages} {315} (\bibinfo {year} {2019})}\BibitemShut {NoStop}%
\bibitem [{\citenamefont {Sinchenko}\ \emph {et~al.}(2022)\citenamefont {Sinchenko}, \citenamefont {Ballou}, \citenamefont {Lorenzo}, \citenamefont {Grenet},\ and\ \citenamefont {Monceau}}]{Sinchenko}%
  \BibitemOpen
  \bibfield  {author} {\bibinfo {author} {\bibfnamefont {A.~A.}\ \bibnamefont {Sinchenko}}, \bibinfo {author} {\bibfnamefont {R.}~\bibnamefont {Ballou}}, \bibinfo {author} {\bibfnamefont {J.~E.}\ \bibnamefont {Lorenzo}}, \bibinfo {author} {\bibfnamefont {T.}~\bibnamefont {Grenet}},\ and\ \bibinfo {author} {\bibfnamefont {P.}~\bibnamefont {Monceau}},\ }\bibfield  {title} {\bibinfo {title} {Does {$(\mathrm{TaSe}_{4})_{2}\mathrm{I}$} really harbor an axionic charge density wave?},\ }\href {https://doi.org/10.1063/5.0080380} {\bibfield  {journal} {\bibinfo  {journal} {Applied Physics Letters}\ }\textbf {\bibinfo {volume} {120}},\ \bibinfo {pages} {063102} (\bibinfo {year} {2022})}\BibitemShut {NoStop}%
\bibitem [{\citenamefont {Sun}\ \emph {et~al.}(2022)\citenamefont {Sun}, \citenamefont {Zhao}, \citenamefont {Palomaki}, \citenamefont {Fei}, \citenamefont {Runburg}, \citenamefont {Malinowski}, \citenamefont {Huang}, \citenamefont {Cenker}, \citenamefont {Cui}, \citenamefont {Chu} \emph {et~al.}}]{sun2022evidence}%
  \BibitemOpen
  \bibfield  {author} {\bibinfo {author} {\bibfnamefont {B.}~\bibnamefont {Sun}}, \bibinfo {author} {\bibfnamefont {W.}~\bibnamefont {Zhao}}, \bibinfo {author} {\bibfnamefont {T.}~\bibnamefont {Palomaki}}, \bibinfo {author} {\bibfnamefont {Z.}~\bibnamefont {Fei}}, \bibinfo {author} {\bibfnamefont {E.}~\bibnamefont {Runburg}}, \bibinfo {author} {\bibfnamefont {P.}~\bibnamefont {Malinowski}}, \bibinfo {author} {\bibfnamefont {X.}~\bibnamefont {Huang}}, \bibinfo {author} {\bibfnamefont {J.}~\bibnamefont {Cenker}}, \bibinfo {author} {\bibfnamefont {Y.-T.}\ \bibnamefont {Cui}}, \bibinfo {author} {\bibfnamefont {J.-H.}\ \bibnamefont {Chu}}, \emph {et~al.},\ }\bibfield  {title} {\bibinfo {title} {Evidence for equilibrium exciton condensation in monolayer {$\mathrm{W} \mathrm{Te_{2}}$}},\ }\href {https://doi.org/10.1038/s41567-021-01427-5} {\bibfield  {journal} {\bibinfo  {journal} {Nature Physics}\ }\textbf {\bibinfo {volume} {18}},\ \bibinfo {pages} {94} (\bibinfo {year} {2022})}\BibitemShut {NoStop}%
\bibitem [{\citenamefont {Fei}\ \emph {et~al.}(2017)\citenamefont {Fei}, \citenamefont {Palomaki}, \citenamefont {Wu}, \citenamefont {Zhao}, \citenamefont {Cai}, \citenamefont {Sun}, \citenamefont {Nguyen}, \citenamefont {Finney}, \citenamefont {Xu},\ and\ \citenamefont {Cobden}}]{fei2017edge}%
  \BibitemOpen
  \bibfield  {author} {\bibinfo {author} {\bibfnamefont {Z.}~\bibnamefont {Fei}}, \bibinfo {author} {\bibfnamefont {T.}~\bibnamefont {Palomaki}}, \bibinfo {author} {\bibfnamefont {S.}~\bibnamefont {Wu}}, \bibinfo {author} {\bibfnamefont {W.}~\bibnamefont {Zhao}}, \bibinfo {author} {\bibfnamefont {X.}~\bibnamefont {Cai}}, \bibinfo {author} {\bibfnamefont {B.}~\bibnamefont {Sun}}, \bibinfo {author} {\bibfnamefont {P.}~\bibnamefont {Nguyen}}, \bibinfo {author} {\bibfnamefont {J.}~\bibnamefont {Finney}}, \bibinfo {author} {\bibfnamefont {X.}~\bibnamefont {Xu}},\ and\ \bibinfo {author} {\bibfnamefont {D.~H.}\ \bibnamefont {Cobden}},\ }\bibfield  {title} {\bibinfo {title} {Edge conduction in monolayer {$\mathrm{W} \mathrm{Te_{2}}$}},\ }\href {https://doi.org/10.1038/nphys4091} {\bibfield  {journal} {\bibinfo  {journal} {Nature Physics}\ }\textbf {\bibinfo {volume} {13}},\ \bibinfo {pages} {677} (\bibinfo {year} {2017})}\BibitemShut {NoStop}%
\bibitem [{\citenamefont {Nielsen}\ and\ \citenamefont {Ninomiya}(1981{\natexlab{a}})}]{nielsen1981absence1}%
  \BibitemOpen
  \bibfield  {author} {\bibinfo {author} {\bibfnamefont {H.~B.}\ \bibnamefont {Nielsen}}\ and\ \bibinfo {author} {\bibfnamefont {M.}~\bibnamefont {Ninomiya}},\ }\bibfield  {title} {\bibinfo {title} {Absence of neutrinos on a lattice: {$(I)$}. {Proof by homotopy theory}},\ }\href {https://doi.org/https://doi.org/10.1016/0550-3213(81)90361-8} {\bibfield  {journal} {\bibinfo  {journal} {Nuclear Physics B}\ }\textbf {\bibinfo {volume} {185}},\ \bibinfo {pages} {20} (\bibinfo {year} {1981}{\natexlab{a}})}\BibitemShut {NoStop}%
\bibitem [{\citenamefont {Nielsen}\ and\ \citenamefont {Ninomiya}(1981{\natexlab{b}})}]{nielsen1981absence2}%
  \BibitemOpen
  \bibfield  {author} {\bibinfo {author} {\bibfnamefont {H.~B.}\ \bibnamefont {Nielsen}}\ and\ \bibinfo {author} {\bibfnamefont {M.}~\bibnamefont {Ninomiya}},\ }\bibfield  {title} {\bibinfo {title} {Absence of neutrinos on a lattice: {$(II)$}. {Intuitive topological proof}},\ }\href {https://doi.org/https://doi.org/10.1016/0550-3213(81)90524-1} {\bibfield  {journal} {\bibinfo  {journal} {Nuclear Physics B}\ }\textbf {\bibinfo {volume} {193}},\ \bibinfo {pages} {173} (\bibinfo {year} {1981}{\natexlab{b}})}\BibitemShut {NoStop}%
\bibitem [{\citenamefont {Wei}\ \emph {et~al.}(2012)\citenamefont {Wei}, \citenamefont {Chao},\ and\ \citenamefont {Aji}}]{Wei2012}%
  \BibitemOpen
  \bibfield  {author} {\bibinfo {author} {\bibfnamefont {H.}~\bibnamefont {Wei}}, \bibinfo {author} {\bibfnamefont {S.-P.}\ \bibnamefont {Chao}},\ and\ \bibinfo {author} {\bibfnamefont {V.}~\bibnamefont {Aji}},\ }\bibfield  {title} {\bibinfo {title} {Excitonic phases from {Weyl} semimetals},\ }\href {https://doi.org/10.1103/PhysRevLett.109.196403} {\bibfield  {journal} {\bibinfo  {journal} {Phys. Rev. Lett.}\ }\textbf {\bibinfo {volume} {109}},\ \bibinfo {pages} {196403} (\bibinfo {year} {2012})}\BibitemShut {NoStop}%
\bibitem [{Note1()}]{Note1}%
  \BibitemOpen
  \bibinfo {note} {We consider the simplest situation, where the band-bending effects near the surface are neglected. In the WSM phase band-bending creates a spiraling structure of the Fermi arcs~\cite {li2015spiraling}. The influence of band-bending on the surface spectrum in the CDW phase is outside the scope of the present work.}\BibitemShut {Stop}%
\bibitem [{\citenamefont {Potter}\ \emph {et~al.}(2014)\citenamefont {Potter}, \citenamefont {Kimchi},\ and\ \citenamefont {Vishwanath}}]{Potter2014}%
  \BibitemOpen
  \bibfield  {author} {\bibinfo {author} {\bibfnamefont {A.~C.}\ \bibnamefont {Potter}}, \bibinfo {author} {\bibfnamefont {I.}~\bibnamefont {Kimchi}},\ and\ \bibinfo {author} {\bibfnamefont {A.}~\bibnamefont {Vishwanath}},\ }\bibfield  {title} {\bibinfo {title} {Quantum oscillations from surface {Fermi arcs in Weyl and Dirac semimetals}},\ }\href {https://doi.org/10.1038/ncomms6161} {\bibfield  {journal} {\bibinfo  {journal} {Nature Communications}\ }\textbf {\bibinfo {volume} {5}},\ \bibinfo {pages} {5161} (\bibinfo {year} {2014})}\BibitemShut {NoStop}%
\bibitem [{\citenamefont {Moll}\ \emph {et~al.}(2016)\citenamefont {Moll}, \citenamefont {Nair}, \citenamefont {Helm}, \citenamefont {Potter}, \citenamefont {Kimchi}, \citenamefont {Vishwanath},\ and\ \citenamefont {Analytis}}]{moll2016transport}%
  \BibitemOpen
  \bibfield  {author} {\bibinfo {author} {\bibfnamefont {P.~J.}\ \bibnamefont {Moll}}, \bibinfo {author} {\bibfnamefont {N.~L.}\ \bibnamefont {Nair}}, \bibinfo {author} {\bibfnamefont {T.}~\bibnamefont {Helm}}, \bibinfo {author} {\bibfnamefont {A.~C.}\ \bibnamefont {Potter}}, \bibinfo {author} {\bibfnamefont {I.}~\bibnamefont {Kimchi}}, \bibinfo {author} {\bibfnamefont {A.}~\bibnamefont {Vishwanath}},\ and\ \bibinfo {author} {\bibfnamefont {J.~G.}\ \bibnamefont {Analytis}},\ }\bibfield  {title} {\bibinfo {title} {Transport evidence for {Fermi-arc-mediated} chirality transfer in the {Dirac} semimetal {$\mathrm{Cd}_{3} \mathrm{As}_{2}$}},\ }\href {https://doi.org/10.1038/nature18276} {\bibfield  {journal} {\bibinfo  {journal} {Nature}\ }\textbf {\bibinfo {volume} {535}},\ \bibinfo {pages} {266} (\bibinfo {year} {2016})}\BibitemShut {NoStop}%
\bibitem [{\citenamefont {Analytis}\ \emph {et~al.}(2010)\citenamefont {Analytis}, \citenamefont {McDonald}, \citenamefont {Riggs}, \citenamefont {Chu}, \citenamefont {Boebinger},\ and\ \citenamefont {Fisher}}]{Analytis2010}%
  \BibitemOpen
  \bibfield  {author} {\bibinfo {author} {\bibfnamefont {J.~G.}\ \bibnamefont {Analytis}}, \bibinfo {author} {\bibfnamefont {R.~D.}\ \bibnamefont {McDonald}}, \bibinfo {author} {\bibfnamefont {S.~C.}\ \bibnamefont {Riggs}}, \bibinfo {author} {\bibfnamefont {J.-H.}\ \bibnamefont {Chu}}, \bibinfo {author} {\bibfnamefont {G.~S.}\ \bibnamefont {Boebinger}},\ and\ \bibinfo {author} {\bibfnamefont {I.~R.}\ \bibnamefont {Fisher}},\ }\bibfield  {title} {\bibinfo {title} {Two-dimensional surface state in the quantum limit of a topological insulator},\ }\href {https://doi.org/10.1038/nphys1861} {\bibfield  {journal} {\bibinfo  {journal} {Nature Physics}\ }\textbf {\bibinfo {volume} {6}},\ \bibinfo {pages} {960} (\bibinfo {year} {2010})}\BibitemShut {NoStop}%
\bibitem [{\citenamefont {Abrikosov}(1988)}]{Abrikosov-book}%
  \BibitemOpen
  \bibfield  {author} {\bibinfo {author} {\bibfnamefont {A.~A.}\ \bibnamefont {Abrikosov}},\ }\href@noop {} {\emph {\bibinfo {title} {Fundamentals of the Theory of Metals}}}\ (\bibinfo  {publisher} {North-Holland, Amsterdam},\ \bibinfo {year} {1988})\BibitemShut {NoStop}%
\bibitem [{\citenamefont {Ashcroft}\ and\ \citenamefont {Mermin}(1976)}]{Ashcroft}%
  \BibitemOpen
  \bibfield  {author} {\bibinfo {author} {\bibfnamefont {N.~W.}\ \bibnamefont {Ashcroft}}\ and\ \bibinfo {author} {\bibfnamefont {N.~D.}\ \bibnamefont {Mermin}},\ }\href@noop {} {\emph {\bibinfo {title} {Solid state physics}}}\ (\bibinfo  {publisher} {Saunders College Publishing},\ \bibinfo {year} {1976})\BibitemShut {NoStop}%
\bibitem [{\citenamefont {Li}\ and\ \citenamefont {Andreev}(2015)}]{li2015spiraling}%
  \BibitemOpen
  \bibfield  {author} {\bibinfo {author} {\bibfnamefont {S.}~\bibnamefont {Li}}\ and\ \bibinfo {author} {\bibfnamefont {A.}~\bibnamefont {Andreev}},\ }\bibfield  {title} {\bibinfo {title} {Spiraling fermi arcs in weyl materials},\ }\href {https://doi.org/10.1103/PhysRevB.92.201107} {\bibfield  {journal} {\bibinfo  {journal} {Physical Review B}\ }\textbf {\bibinfo {volume} {92}},\ \bibinfo {pages} {201107} (\bibinfo {year} {2015})}\BibitemShut {NoStop}%
\end{thebibliography}%

\end{document}